\documentclass[amsmath, amssymb, aps, showpacs, prd, twocolumn,groupedaddress]{revtex4-1}
\usepackage[T1]{fontenc}
\usepackage{ae,aecompl}

\usepackage{enumerate}
\usepackage[colorlinks=false, linkcolor=black, citecolor=black]{hyperref}
\usepackage{graphicx}
\usepackage{subfig}

\newcommand{\rd}{\mathrm{d}}
\newcommand{\re}{\mathrm{e}}
\newcommand{\Oin}{\Omega^{\text{(in)}}}
\newcommand{\Oout}{\Omega^{\text{(out)}}}

\begin{document}

\title{Interpretation of the Weyl tensor}

\author{Stefan Hofmann}
\email[]{stefan.hofmann@physik.uni-muenchen.de}
\author{Florian Niedermann}
\email[]{florian.niedermann@physik.uni-muenchen.de}
\author{Robert Schneider}
\email[]{robert.bob.schneider@physik.uni-muenchen.de}
\affiliation{Arnold Sommerfeld Center for Theoretical Physics, Ludwig-Maximilians-Universit\"at, Theresienstra{\ss}e 37, 80333 Munich, Germany}
\affiliation{Excellence Cluster Universe, Boltzmannstra{\ss}e 2, 85748 Garching, Germany}

\date{\today}

\begin{abstract}

According to folklore in general relativity, the Weyl tensor can be decomposed 
into parts corresponding to Newton-like, incoming and outgoing wavelike field components. 
It is shown here that this one-to-one correspondence does not hold for space-time geometries with cylindrical isometries.
This is done by investigating some well-known exact solutions of Einstein's field equations with whole-cylindrical symmetry, for which the physical interpretation is very clear, but for which the standard Weyl interpretation would give contradictory results.
For planar or spherical geometries, however, the standard interpretation
works for both, static and dynamical space-times.
It is argued that one reason for the failure in the cylindrical case is that for waves spreading in two spatial dimensions there is no local criterion to distinguish incoming and outgoing waves already at the linear level.
It turns out that Thorne's local energy notion, subject to certain qualifications, provides an efficient diagnostic tool to extract the proper physical interpretation of the space-time geometry in the case of cylindrical configurations.

\end{abstract}

\pacs{04.20.-q, 04.30.-w, 04.20.Jb}

\maketitle

\section{Introduction}

Newton's theory of a gravitational force is given by the pair $(\mathcal{G},\Phi)$, where 
$\mathcal{G}$ denotes Galilean space and $\Phi$ is the gravitational potential.
In Einstein's theory of general relativity, the pair $(\mathcal{G},\Phi)$ is superseded 
by $(\mathcal{M},g)$, where $\mathcal{M}$ denotes space-time, modeled as a pseudo-Riemannian
manifold with a local geometry represented by a metric field $g$. 
While in the classical theory the interpretation of $ \Phi $ is completely clear --- its gradient simply gives (minus) the acceleration that a test particle would undergo --- an analogous interpretation of $ g $ and its derivatives is far from obvious in the relativistic theory.
This is mainly due to the equivalence principle, saying that the affine connection (which is basically the first derivative of the metric, and thus analogue to the Newtonian force field) can be set to zero locally at any point $ p \in \mathcal{M} $ by an appropriate choice of coordinates. A real gravitational effect is only present at $ p $, if this cannot be done in a whole neighborhood of $ p $ simultaneously, or equivalently if the Riemann tensor at $ p $ does not vanish.
Outside sources, the Riemann tensor reduces to the Weyl tensor $C$. Thus, the physical content of the metric field in vacuum should be somehow encoded in $C$. Moreover, instead of having only one static field component $ \Phi $, there are new dynamical degrees of freedom, corresponding to gravitational waves. It would thus be desirable to have a decomposition of the Weyl tensor into Newton-like parts and wavelike parts.

Since Einstein's field equations are nonlinear and coupled, the possibility of a clear separation between 
Newton-like and wavelike contributions might be doubtful. 
However, Szekeres \cite{Szekeres:1965ux} showed that such a decomposition can in fact be found, by following Pirani's approach \cite{Pirani:1956} and studying the geodesic deviation for nearby freely falling test particles. More recently, this program has been generalized by Podolsky and Svarc \cite{Podolsky:2012he}
to space-times of arbitrary dimensions.

The main result of these investigations is that $C$ can be deconstructed into components
corresponding to Newton-like\footnote{We will use the more appropriate name ``Newton-like'', instead of Szekeres' ``Coulomb-like'' \cite {Szekeres:1965ux}, which has become standard terminology. Some authors also choose the compromise ``Newton-Coulomb-like'' \cite {Podolsky:2012he}.} components and, in addition, to contributions
corresponding to transverse and longitudinal, outgoing and incoming gravitational waves.
We will refer to this result as the \textit{standard interpretation} of the Weyl tensor.
The standard interpretation is frequently quoted and used in the literature, 
see e.\,g.~\cite{Stephani, Wang:1991nf, Beetle:2002, Nolan:2004, Nerozzi:2006aj, Nerozzi:2008ng}.

In this paper we will show that, although the standard interpretation works very well in certain cases, there are other cases where it fails. We will do so by studying two explicit examples of exact solutions of Einstein's field equations, for which the true physical interpretation is evident, but where the standard interpretation would give contradictory results.
To be more specific, we show that a nonzero component of the Weyl tensor does not imply that the corresponding physical effect is present. The correct physical interpretation can, however, still be deduced from the asymptotic falloff behavior of those components.

The paper is organized as follows:
We start by fixing some conventions in section \ref{sec:conventions}. Then, in section \ref{sec:interpret_weyl} we review the standard interpretation of the Weyl tensor, introduced by Pirani \cite{Pirani:1956, Pirani:2009} and Szekeres \cite{Szekeres:1965ux} using the 
concept of geodesic deviation.
Section \ref{sec:interpret_works} is devoted to the discussion of planar and spherical geometries, 
both static and dynamical, for which the interpretation works very well. 
Then, in section \ref{sec:interpret_fails} we turn to solutions of Einstein's equations 
with cylindrical isometries, again static as well as dynamical, and show that the standard interpretation breaks down. In section \ref{sec:pirani} we note that the static cylindrically symmetric solution also invalidates Pirani's wave criterion \cite{Pirani:1957}, and in section \ref{sec:thorne} we check that Thorne's C-energy \cite{Thorne:1965} provides a good tool for cylindrical setups. Finally, we conclude in section \ref{sec:conclusion}.

\section{Conventions\label{sec:conventions}}

Our notational conventions are the following:
capital latin indices $ A, B, \dots $ denote $ d $-dimensional space-time indices (in the examples considered,
$ d = 4 $), boldface symbols denote vector-fields, and small latin indices $ a, b, \ldots $ 
are tetrad indices. Small latin indices $ i, j, \ldots $ are those tetrad indices which run only over 
the $ d-2 $ spatial tetrad vectors orthogonal to the privileged spatial direction (see below), 
whereas tetrad indices evaluated as $ 0 $ or $ 1 $ always correspond to the two null 
tetrad vectors, cf.\ equation \eqref{eq:def_mixed_tet}. Our sign conventions are ``$ +++ $'' as defined (and adopted) in \cite{Misner}. Antisymmetrization of indices is denoted by square brackets, i.e.
\begin{equation}
	T_{[AB]} := \frac{1}{2} \left( T_{AB} - T_{BA} \right),
\end{equation}
and we also use the notation
\begin{equation}
	T_{\{ABCD\}} := \frac{1}{2} \left ( T_{[AB][CD]} + T_{[CD][AB]} \right ).
\end{equation}
We work in units in which $ c = 8\pi G = 1 $.

\section{Standard Interpretation of Weyl components\label{sec:interpret_weyl}}

Let us first briefly review the technique originally used by Szekeres \cite{Szekeres:1965ux} to extract the physical meaning of the various components of the Weyl tensor, before we apply it to some concrete solutions of Einstein's field equations.

Consider a timelike geodesic with unit tangent vector (field) $ \mathbf{t} $, and a neighboring timelike geodesic separated by the vector $ \mathbf{y} $ (parametrized such that $ \mathbf{t} \cdot \mathbf{y} = 0 $). Then, the geodesic deviation equation, governing the change of $ \mathbf{y} $ at linear order, in vacuum reads
\begin{equation}
\label{eq:geodesic_dev}
	\frac{\mathrm{D}^2 y^A}{\mathrm{d}\tau^2} = {C^A}_{BCD}\, t^B t^C y^D =: {M^A}_D\, y^D \; ,
\end{equation}
where $ \frac{\mathrm{D}}{\mathrm{d}\tau} $ denotes covariant differentiation along the geodesic and $ {C^A}_{BCD} $ is the Weyl tensor, i.e.\ the traceless part of the Riemann tensor:
\begin{equation}
\begin{split}
	C_{ABCD} & = R_{ABCD} - \frac{2}{d-2} \left( R_{A[C} g_{B]D} - R_{B[C} g_{D]A} \right) \\
	& \quad + \frac{2}{(d-1)(d-2)} R\, g_{A[C}g_{D]B} \; .
\end{split}
\end{equation}
Now complement $ \mathbf{t} $ with $ d-1 $ orthonormal spacelike vectors $ \mathbf{x}_1, \ldots \mathbf{x}_{d-1} $ to obtain an orthonormal tetrad
\begin{equation}
	\mathbf{e}_{a} = \left ( \mathbf{t},\, \mathbf{x}_1,\, \ldots\, , \mathbf{x}_{d-1} \right) ,\quad \mathbf{e}_a \cdot \mathbf{e}_b = \eta_{ab}\;.
\end{equation}
This tetrad defines a frame which an observer along the geodesic would use to make physical measurements. It is further convenient to define a null tetrad, but instead of working with a complex null tetrad, as is usually done in $ d = 4 $, we will use the conventions of Podolsky and Svarc \cite{Podolsky:2012he} and take a real null tetrad, or \textit{mixed tetrad}, since this approach has the advantage of being applicable in any number of dimensions. The complexification of the two remaining spatial dimensions is in fact of no importance for the discussion, and the results will of course be completely equivalent. Therefore, we define
\begin{equation}
\label{eq:def_mixed_tet}
	\mathbf{m}_a = (\mathbf{m}_0,\, \mathbf{m}_1,\, \mathbf{m}_i) := (\mathbf{k},\, \mathbf{l},\, \mathbf{m}_i) \quad \left(i = 2, \ldots, d-1\right),
\end{equation}
by combining one of the spacelike vectors, say $ \mathbf{x}_1 =: \mathbf{x} $, with $ \mathbf{t} $ to form two appropriately normalized null vectors, and leaving the other vectors unchanged:
\begin{subequations}
\begin{gather}
	\mathbf{k} := \frac{1}{\sqrt{2}} \left ( \mathbf{t} + \mathbf{x} \right ), \qquad \mathbf{l} := \frac{1}{\sqrt{2}} \left ( \mathbf{t} - \mathbf{x} \right ),\\
	\qquad \mathbf{m}_i := \mathbf{x}_i \quad \left(i = 2, \ldots, d-1\right).
\end{gather}
\end{subequations}
This frame now satisfies the quasiorthonormality relations
\begin{subequations}
	\begin{align}
		\mathbf{k}\cdot\mathbf{k} = \mathbf{l}\cdot\mathbf{l} &= 0 \; , & \mathbf{k} \cdot \mathbf{l} &= -1 \; , \\
		\mathbf{m}_i \cdot \mathbf{k} = \mathbf{m}_i \cdot \mathbf{l} &= 0 \; , & \mathbf{m}_i\cdot\mathbf{m}_j &= \delta_{ij} \; ,
	\end{align}
\end{subequations}
or, in matrix notation
\begin{equation}
	\textbf{m}_a \cdot \mathbf{m}_b = \tilde{\eta}_{ab} :=
	\begin{pmatrix}
	0 & -1 & 0 \\
	-1 & 0 & 0 \\
	0 & 0 & \delta_{ij}
	\end{pmatrix},
\end{equation}
and the metric can be written as
\begin{equation}
	g_{AB} = - k_A l_B - l_A k_B + m^2_A m^2_B + \ldots + m^{d-1}_A m^{d-1}_B \; .
\end{equation}
Note that here, and in what follows, all evaluated indices (and indices $ i, j, \ldots $) are to be understood as mixed tetrad indices, as defined in \eqref{eq:def_mixed_tet}. Space-time indices will always remain unevaluated and denoted by $ A, B, \ldots $. The mixed tetrad indices are raised and lowered with $ \tilde{\eta}_{ab} $ (and its inverse), in particular
\begin{equation}
	\mathbf{m}^a = \left( \mathbf{m}^0,\, \mathbf{m}^1,\, \mathbf{m}^i \right) = \left( -\mathbf{l},\, -\mathbf{k},\, \mathbf{m}_i \right).
\end{equation}

The Weyl tensor can now be expressed in terms of its mixed tetrad components
\begin{equation}
	C_{ABCD} = C_{abcd}\, m^a_A m^b_B m^c_C m^d_D \; ,
\end{equation}
which, using its symmetries $ C_{ABCD} = C_{\{ABCD\}} $, can be expanded as
\begin{equation}
\label{eq:weyl_decomp}
	\begin{split}
		& C_{ABCD} = \ 4 C_{0i0j}\, l_{\{A} m^i_B l_C m^j_{D\}} \\
		&\quad - 8 C_{010i}\, l_{\{A} k_B l_C m^i_{D\}} - 4 C_{0ijk}\, l_{\{A} m^i_B m^j_C m^k_{D\}} \\
		&\quad + 4 C_{0101}\, l_{\{A} k_B l_C k_{D\}} + 4 C_{01ij}\, l_{\{A} k_B m^i_C m^j_{D\}} \\
		&\quad + 8 C_{0i1j}\, l_{\{A} m^i_B k_C m^j_{D\}} + C_{ijkl}\, m^i_{\{A} m^j_B m^k_C m^l_{D\}} \\
		&\quad - 8 C_{101i}\, k_{\{A} l_B k_C m^i_{D\}} - 4 C_{1ijk}\, k_{\{A} m^i_B m^j_C m^k_{D\}} \\
		&\quad + 4 C_{1i1j}\, k_{\{A} m^i_B k_C m^j_{D\}} \; .
	\end{split}
\end{equation}
Here, the terms have been ordered according to their \textit{boost weights} \cite{Coley:2004jv}, where some quantity $ Q $ is said to have boost weight $ b $, if it transforms under a Lorentz boost with velocity $ v $ in direction $ \mathbf{x} $ as
\begin{align}
	& Q \mapsto \lambda^{b} Q \; , && \lambda := \sqrt{\frac{1+v}{1-v}} > 1 \; .
\end{align}
The physical interpretation of each term is then found by inserting \eqref{eq:weyl_decomp} into \eqref{eq:geodesic_dev} and using the various orthogonality relations. The corresponding contributions to the matrix $ M_{AD} $ in \eqref{eq:geodesic_dev} are (the notation follows \cite{Durkee:2010xq}):
\begin{enumerate}[(i)]
	\item\underline{\textit{boost weight} $ \mathit{+2} $:}
	\begin{equation}
	C_{0i0j} =: \Omega_{ij}  \longrightarrow -\frac{1}{2} \Omega_{ij} m^i_A m^j_D
	\end{equation}
	This gives rise to a deviation $ y^D $ of the neighboring geodesic into the hyperplane\footnote{We use the terms ``hyperplane'', ``sphere'' and ``ellipsoid'' in order to include higher-dimensional space-times. In $ d = 4 $, these terms can be replaced by ``plane'', ``circle'' and ``ellipse''.} spanned by the $ \mathbf{m}^i $, i.e.\ orthogonal to $ \mathbf{x} $. Furthermore, since the matrix $ \Omega_{ij} $ is symmetric and traceless, a sphere of test particles in this hyperplane will be deformed into an ellipsoid. This is the characteristic effect of a gravitational wave, and so $ \Omega_{ij} $ is usually interpreted as (the modes of) a \emph{transverse gravitational wave propagating in the direction} $ -\mathbf{x} $. The direction of propagation can for instance be inferred from the boost weight, which is $ +2 $ for $ \Omega_{ij} $. This means that a Lorentz boost in the direction $ +\mathbf{x} $ will enhance this term, in accordance with the interpretation of a wave traveling in the opposite direction. Alternatively, inserting a plane wave traveling in direction $ +\mathbf{x} $ shows that this term actually vanishes; cf.\ section \ref{sec:plane_wave}.

	\item\underline{\textit{boost weight} $ \mathit{+1} $:}
	\begin{equation}
		C_{010i} =: \Psi_i \longrightarrow \frac{1}{\sqrt{2}} \Psi_i \left( x_A m^i_D + m^i_A x_D \right) 
	\end{equation}
	This term is similar to the first one, but this time the deflection occurs in the $ \mathbf{x} $-$ \mathbf{m}^i $ hyperplane. Therefore, it is usually identified as a \textit{longitudinal gravitational wave propagating in the direction} $ -\mathbf{x} $.
	
	\item\underline{\textit{boost weight} $ \mathit{0} $:}
	\begin{multline}
	C_{0101} =: \Phi \; , \; C_{0i1j} =: \Phi_{ij} \\
	\longrightarrow - \left( \Phi x_A x_D + \Phi_{(ij)} m^i_A m^j_D \right)
	\end{multline}
	Due to the traceless condition, $ \Phi + {\Phi^i}_i = 0 $, this term will stretch a sphere of test particles in the $ \mathbf{x} $ direction, while leading to contractions in the directions $ \mathbf{m^i} \perp \mathbf{x}$ (or vice versa). Thus, it represents the (higher-dimensional) analogue of the tidal forces caused by localized, static sources, and can, therefore, be interpreted as a \textit{Newton-like} part of the Weyl tensor. 

	\item\underline{\textit{boost weight} $ \mathit{ -1} $:}
	\begin{equation}
		C_{101i} =: \Psi'_i \longrightarrow \frac{1}{\sqrt{2}} \Psi'_i \left( x_A m^i_D + m^i_A x_D \right)
	\end{equation}
	In complete analogy to case (ii), this term is interpreted as a \textit{longitudinal gravitational wave propagating in the direction} $ +\mathbf{x} $.

	\item\underline{\textit{boost weight} $ \mathit{ -2} $:}
	\begin{equation}
		C_{1i1j} =: \Omega'_{ij} \longrightarrow -\frac{1}{2} \Omega'_{ij} m^i_A m^j_D
	\end{equation}

	In analogy to case (i), this term is interpreted as a \textit{transverse gravitational wave propagating in the direction} $ +\mathbf{x} $.
	
\end{enumerate}

All other terms, i.e.\ those standing on the right in \eqref{eq:weyl_decomp}, only give vanishing contributions to \eqref{eq:geodesic_dev}. This does by no means imply that these components of the Weyl tensor have no physical effect, it only implies that they give no linear order contribution to the relative acceleration of freely falling nearby test particles.

The privileged spatial vector $ \mathbf{x} $, which is the direction of propagation of the gravitational wave components, can be chosen arbitrarily. In the general case it is of course possible to have superpositions of waves traveling in any direction. The geometries considered by us explicitly, however, have 
sufficient isometries to single out a unique direction of propagation.
This means that the two null vectors $ \mathbf{k} $ and $ \mathbf{l} $ are uniquely determined, apart from Lorentz boosts in the direction of wave propagation, a point which we will come back to later. There is also some freedom in choosing the remaining orthogonal tetrad vectors $ \mathbf{m}_i $, corresponding to rotations in the plane orthogonal to the wave direction. This will however not affect any of our results.

\section{Examples where the standard interpretation works\label{sec:interpret_works}}

Let us first look at some exact vacuum solutions of Einstein's field equations, which support the standard interpretation of the Weyl components.

\subsection{Static plane\label{sec:static_plane}}

The general plane-symmetric, static solution of Einstein's field equations in vacuum was first found by Levi-Civita \cite{Levi:1918} (see also \cite{Amundsen:1983} for a more recent discussion), and can be written in the form
\begin{subequations}
\label{eq:met_static_plane}
\begin{align}
	\rd s^2 & = - a^{-2/3} \rd t^2 + \rd x^2 + a^{4/3}  \left(\rd y^2 + \rd z^2 \right),\\
	a(x) & = 1 - \alpha x \; ,
\end{align}\end{subequations}
where $ \alpha $ is some constant, and we used the gauge freedom to set $ a(x=0) = 1 $. As a  source which gives rise to this geometry, we consider a thin layer of matter located at $ x = 0 $, and so according to the symmetries of the set-up, the energy-momentum tensor has the form
\begin{subequations}
\begin{align}
\label{eq:emt_static_plane}
	T^{A}_{\phantom{A}B} & = \text{diag}\left(-\rho, 0, p, p \right) \delta(x) \; .
\end{align}
\end{subequations}

For $ x \neq 0 $, the metric is of the form \eqref{eq:met_static_plane}.
By choice, it is continuous across $ x = 0 $ and accommodates different constants
$\alpha$ in the half-spaces $ x > 0 $ and $ x < 0 $, say $ \alpha_> $ and $ \alpha_< $.
These constants are not arbitrary, however, and must be chosen such that 
the discontinuity in the first derivative of the metric with respect to $x$
implies, upon taking one more $x$ derivative, 
the correct $ \delta $-like contribution, as given by \eqref{eq:emt_static_plane}.
This procedure corresponds to implementing Israel's junction conditions \cite{Israel:1966, Israel:1967}. 
We will further assume the full metric to be symmetric across the plane, 
i.e.\ under $ x \mapsto -x  $, implying $ \alpha_> = - \alpha_< =: \alpha $. 
Then, the junction conditions yield
\begin{equation}	
	\frac{8 \alpha}{3} = \rho, \quad
	\frac{2 \alpha}{3} = - p
	\quad \Rightarrow p = -\frac{\rho}{4} \; .
\end{equation}
The matter source has to fulfill this equation of state (implying a negative pressure) in order to allow for a static solution. Note that $ \alpha > 0 $ for a physically reasonable source with positive energy density.

The symmetry of this geometry suggests that we choose $ \mathbf{x} = \boldsymbol{\partial}_x $ as the privileged direction, with respect to which we calculate the Weyl components. The corresponding mixed tetrad is
\begin{subequations}
\label{eq:mixed_tet_static_plane}
\begin{gather}
	\mathbf{k} = \frac{1}{\sqrt{2}}\left( a^{1/3} \boldsymbol{\partial}_t + \boldsymbol{\partial}_x\right), \quad
	\mathbf{l} = \frac{1}{\sqrt{2}}\left( a^{1/3} \boldsymbol{\partial}_t - \boldsymbol{\partial}_x\right), \\
	\mathbf{m}_i = a^{-2/3} \boldsymbol{\partial}_i \; , \quad \left( i = y, z \right).
\end{gather}
\end{subequations}
However, there is a problem with this choice of tetrad: for the interpretation discussed above to apply, the timelike vector must be tangent to a geodesic. But $ a^{1/3} \boldsymbol{\partial}_t $ is in fact \textit{not} parallel transported along its integral curves (except for the trivial case $ \alpha = 0 $), so the frame \eqref{eq:mixed_tet_static_plane} can actually not be used. However, since the metric \eqref{eq:met_static_plane} admits the Killing vectors $ \mathbf{m}_i $, it is clear that geodesics exist with tangent vectors of the form $ \mathbf{t} = f \boldsymbol{\partial}_t + g \boldsymbol{\partial}_x $ with some functions $ f(x) $ and $ g(x) $. The corresponding null vectors have the form $ h^{\pm 1} \left( a^{1/3} \boldsymbol{\partial}_t \pm \boldsymbol{\partial}_x \right) $ with some function $ h(x) $, which can be seen by requiring the various orthonormality relations among the vectors to hold. Therefore, the Weyl components in the frame \eqref{eq:mixed_tet_static_plane} only differ from the ones for which the physical interpretation was derived by the overall factors $ h^{\pm b} \neq 0 $, where $ b $ is the corresponding boost weight. Hence, since it is sufficient for our purpose to identify the vanishing components, 
we might as well use the frame \eqref{eq:mixed_tet_static_plane}.

Having established the frame, it is straightforward to compute the various Weyl components. It turns out that all of the wave components vanish identically:
\begin{equation}
	\Omega_{ij} = \Psi_i = 0 = \Omega'_{ij} = \Psi'_i \; ,
\end{equation}
and the only nonzero components are the Newton-like terms
\begin{align}
\label{eq:coul_static_plane}
	&\Phi = \frac{4 \alpha^2}{9 \left(1 - \alpha |x|\right)^2} \; , && \Phi_{ij} = -\frac{1}{2} \Phi\, \delta_{ij} \; .
\end{align}
This result is in perfect agreement with the standard interpretation of the various Weyl components.

For completeness, it should be mentioned that there are also some nonvanishing Weyl components in the nonobservable sector of section \ref{sec:interpret_weyl}. They are again all of boost weight 0 and are given by
\begin{equation}
	C_{ijkl} = -\frac{1}{2} \Phi \left( \delta_{ik}\delta_{jl} - \delta_{jk}\delta_{il} \right),
\end{equation}
yielding no further independent components. Their appearance follows from the Newton-like terms \eqref{eq:coul_schw} by the traceless condition $ {C^{i}}_{jik} = C_{0j1k} + C_{1j0k} = 2\Phi_{(jk)} $. Similar comments will also apply in the following examples, but we will not explicitly reiterate on this.

As an aside, note that there are space-time singularities at $ x = \pm 1/\alpha $, as can be seen by calculating the Kretschmann scalar $ R^{ABCD} R_{ABCD} = 12 \Phi^2 $. However, geodesic motion of massive particles cannot reach them, and photons undergo an infinite redshift along geodesics towards the singularities. As a consequence, the physical accessible space-time only covers the open interval $ x \in (-1/\alpha, 1/\alpha) $ \cite{Amundsen:1983}.

\subsection{Schwarzschild\label{sec:schwarzschild}}

In the static Schwarzschild geometry, we expect that again only the Newton-like field components are present. Let us now verify that this is indeed the case. The metric outside the Schwarzschild radius $ r_s $ is given by
\begin{subequations}
\begin{align}
\label{eq:met_schw}
	\rd s^2 & = - f \rd t^2 + f^{-1} \rd r^2 + r^2 \left[ \rd\theta^2 + \sin(\theta)^2 \rd \phi^2 \right],\\
	f(r) & = 1 - \frac{r_s}{r} \; .
\end{align}
\end{subequations}
This time, the symmetry allows us to identify $ \mathbf{x} \propto \boldsymbol{\partial}_r $ as the privileged spatial direction, and so the corresponding mixed tetrad reads
\begin{subequations}
\label{eq:mixed_tet_schw}
\begin{align}
	\mathbf{k} &= \frac{1}{\sqrt{2}}\left( f^{-1/2} \boldsymbol{\partial}_t + f^{1/2} \boldsymbol{\partial}_r \right),\\
	\mathbf{l} &= \frac{1}{\sqrt{2}}\left( f^{-1/2} \boldsymbol{\partial}_t - f^{1/2} \boldsymbol{\partial}_r \right),\\
	\mathbf{m}_1 &= \frac{1}{r} \boldsymbol{\partial}_{\theta} \; , \quad \mathbf{m}_2 = \frac{1}{r \sin(\theta)} \boldsymbol{\partial}_{\phi} \; .
\end{align}
\end{subequations}
As before, $ f^{-1/2} \boldsymbol{\partial}_t $ is not tangent to a geodesic, but a discussion completely analogous to the corresponding situation in the example of a static plane applies, and so we can in fact use this frame. And again, all of the wave components vanish, the only nonzero components being the Newton-like terms
\begin{align}
\label{eq:coul_schw}
	&\Phi = -\frac{r_s}{r^3} \; , && \Phi_{ij} = -\frac{1}{2} \Phi\, \delta_{ij} \; .
\end{align}
As a result, also in the case of a spherically symmetric geometry, the standard interpretation of the Weyl components works perfectly well.

\subsection{Plane wave\label{sec:plane_wave}}

Next, we consider the complementary case: instead of a static field configuration which only has Newton-like components, we investigate a geometry that corresponds to pure plane gravitational waves. In this case, only the wavelike components are expected to be nonzero. Specifically, consider the well-known pp-wave vacuum solution \cite{Brinkmann:1925fr} (see also \citep[chap. 24.5]{Stephani})
\begin{subequations}
\label{eq:met_plane_wave}
\begin{align}
	\rd s^2 & = -2 h \rd u^2 - 2 \rd u \rd v + \rd y^2 + \rd z^2 \; ,\\
	\intertext{with}
	h(u, y, z) & = a(u) \left( y^2 - z^2 \right)/2 + b(u) y z \; ,
\end{align}
\end{subequations}
where 
$ a(u) $ and $ b(u) $ are some arbitrary functions. The exact solution \eqref{eq:met_plane_wave} represents a plane gravitational wave with wave vector $ \boldsymbol{\partial}_v $, and the corresponding mixed tetrad frame reads
\begin{subequations}
\label{eq:mixed_tet_plane_wave}
\begin{gather}
	\mathbf{k} = \boldsymbol{\partial}_v \; , \qquad \mathbf{l} = \boldsymbol{\partial}_u - h \boldsymbol{\partial}_v \; ,\\
	\mathbf{m}_i = \boldsymbol{\partial}_i \; , \quad \left ( i = y, z \right) .
\end{gather}
\end{subequations}
Note that this time, the vector $ \mathbf{k} $ is tangent to a \textit{null} geodesic. However, this poses no conceptual problem, since the standard interpretation of the Weyl tensor as discussed in section \ref{sec:interpret_weyl} can analogously be applied to null geodesics instead of timelike geodesic, cf.\ \cite{Wang:1991nf}.

And indeed, in the frame \eqref{eq:mixed_tet_plane_wave} the only nonvanishing Weyl components are
\begin{equation}
	\Omega'_{ij} =
	\begin{pmatrix}
	a(u) & b(u) \\
	b(u) & -a(u)
	\end{pmatrix},
\end{equation}
while, in particular, $ \Omega_{ij} = 0 $. Hence, the standard interpretation of the Weyl tensor correctly identifies the purely wavelike field character, as well as the direction of propagation of the wave. Evidently, the two arbitrary functions $ a $ and $ b $ are precisely the two modes of the gravitational wave.

\subsection{Spherical wave\label{sec:spherical_wave}}

Let us finally investigate the Robinson-Trautman space-time \cite{Robinson:1960} (see also \citep[chap. 28]{Stephani}), which can be interpreted as describing spherical (but of course not spherically symmetric) gravitational waves. The metric has the form
\begin{equation}
\label{eq:met_spherical_wave}
	\rd s^2 = -2 H \rd u^2 - 2 \rd u \rd r + \frac{r^2}{p^2} \left( \rd \xi^2 + \rd \eta^2 \right),
\end{equation}
where $ p $ is a function of $ (u, \xi, \eta) $ and
\begin{equation}
	H(u, r, \xi, \eta) = \frac{1}{2} \Delta \ln p - r \left( \ln p \right)' - \frac{m(u)}{r} \; ,
\end{equation}
with $ \Delta := p^2 \left( \partial^2_\xi + \partial^2_\eta \right) $ and the prime denoting differentiation with respect to $ u $. With this choice, the vacuum Einstein equations reduce to the single fourth order differential equation
\begin{equation}
\label{eq:EFE_spherical_wave}
	\Delta\Delta \ln(p) + 12 m \left( \ln p \right)' - 4m' = 0 \; .
\end{equation}
Here $ u $ is a retarded time coordinate and $ \boldsymbol{\partial}_r $ is tangent to a null geodesic. We can therefore choose the mixed tetrad frame
\begin{subequations}
\label{eq:mixed_tet_spherical_wave}
\begin{gather}
	\mathbf{k} = \boldsymbol{\partial}_r \; , \qquad \mathbf{l} = \boldsymbol{\partial}_u - H \boldsymbol{\partial}_r \; ,\\
	\mathbf{m}_i = \frac{p}{r} \boldsymbol{\partial}_i \quad \left( i = \xi,\eta \right) ,
\end{gather}
\end{subequations}
in which the nonzero, independent components of the Weyl tensor become
\begin{subequations}
\label{eq:weyl_spherical_wave}
\begin{align}
		\Phi_{ij} & = \frac{m}{r^3} \delta_{ij}\\
		\Psi'_{i} &= -\frac{p}{2r^2} \partial_i \Delta \ln p \\
		\Omega'_{ij} & = \frac{1}{2r}
		\begin{pmatrix}
			A_{11} - A_{22} & A_{12} + A_{21} \\
			A_{21} + A_{12} & -A_{11} + A_{22}
		\end{pmatrix}
		\intertext{with}
		A_{ij} & := \partial_i \left\{ p^2 \partial_j \left[ \frac{1}{2} \Delta \ln p - r \left( \ln p \right)' \right] \right\}.
\end{align}
\end{subequations}
The main result is that indeed all the incoming (i.e.\ in the direction opposite to $ u $) components $ \Psi_i $ and $ \Omega_{ij} $ vanish, and only Newton-like and outgoing wavelike components are present in general. (Note that ``pure wave'' solutions without Newton-like admixture can be constructed  by setting $ m=0 $ .) Therefore, this example is also in perfect agreement with the standard interpretation.

The explicit form of the Weyl components depends on the particular solution of \eqref{eq:EFE_spherical_wave}, one example being \cite{Robinson:1960}
\begin{align}
	m(u) = m_0 = \text{const} \; , &&  p(u, \xi, \eta) = \xi^{3/2} \; ,
\end{align}
for which
\begin{align}
		\Phi_{ij} = \frac{m_0}{r^3} \delta_{ij} \; , &&
		\Psi'_{\xi} = -\frac{3 \xi^{3/2}}{4 r^2} \; , &&
		\Omega'_{ij} = \frac{9 \xi^2}{8 r^2}
		\begin{pmatrix}
			-1 & 0 \\
			0 & 1
		\end{pmatrix}.
\end{align}
These terms have the correct falloff behavior $ \sim 1/r^3 $ and $ \sim 1/r^2 $ for Newton-like and wavelike components, respectively, expected in a spatially three-dimensional, spherical setup\footnote{Note that the Weyl tensor is morally the second derivative of the metric, so the Newtonian $ \sim 1/r $ potential of a spherical symmetric source in three space dimensions would manifest itself as a $ \sim 1/r^3 $ contribution to the Weyl tensor.}.
The scaling $ \sim 1/r^2 $ of the wave components is also in agreement with the asymptotic criterion for outgoing waves put forward in \cite{ADM:1961}.

\section{Examples where the standard interpretation fails\label{sec:interpret_fails}}

In this section we prove that the standard interpretation of the Weyl tensor 
does not hold for all space-time geometries. 
The focus will be on geometries with whole-cylinder symmetry, 
i.e.\ those with azimuthal $ \phi $ symmetry and symmetry along a $ z $ direction 
perpendicular to $ \phi $, for which the metric can most conveniently be written in the form (see e.g.\ \cite{Thorne:1965} or \citep[chap. 22]{Stephani})
\begin{equation}
\label{eq:met_cyl_symm}
	\rd s^2 = \re^{2(\eta - \alpha)} \left( -\rd t^2 + \rd r^2 \right) + \re^{2\alpha} \rd z^2 + \re^{-2\alpha} w^2 \rd\phi^2 \; ,
\end{equation}
where $ \eta, \alpha $ and $ w $ are functions of $ (t, r) $.
Note that \eqref{eq:met_cyl_symm} only describes a cylindrical geometry, if there is an axis (which we will assume to be located at $ r=0 $), on which the norm of the Killing vector associated with $ \phi $ symmetry $\boldsymbol{\sigma} = \boldsymbol{\partial}_{\phi}$ vanishes:
\begin{align} \label{eq:symmetry_axis}
	\sigma^2 \equiv \sigma_{A}\sigma^{A} = w^2 \re^{-2\alpha}  \xrightarrow[]{r \rightarrow 0} 0 \; .
\end{align}
Furthermore, requiring the metric not to have a conical singularity at $ r=0 $ leads to the following \textit{regularity condition} on the axis \cite{Mars:1992cm}:
\begin{multline}\label{eq:regularity_cond}
	\frac{\nabla_{A}(\sigma^2)\nabla^{A}(\sigma^2)}{4\sigma^2} = \\
	= w^2 \re^{-2\eta} \left[ -\left( \frac{\dot w}{w} - \dot\alpha \right)^2 + \left( \frac{w'}{w} - \alpha' \right)^2 \right] \xrightarrow[]{r \rightarrow 0} 1 \; .
\end{multline}
In these coordinates, the Einstein field equations read:
\newcommand{\tT}[2]{\mathcal{T}^{#1}_{\phantom{#1}#2}}
\begin{subequations}
\label{eq:EFE_cyl}
	\begin{align}
		\frac{w''}{w} - \frac{\ddot w}{w} &= \tT{t}{t} + \tT{r}{r} \\
		\alpha'' + \frac{w'}{w} \alpha' - \ddot \alpha - \frac{\dot w}{w} \dot \alpha &= 2\left( \tT{t}{t} + \tT{r}{r} + \tT{\phi}{\phi} - \tT{z}{z}  \right)\\
		{\alpha'}^2 + \dot \alpha^2 - \frac{w'}{w} \eta' - \frac{\dot w}{w} \dot \eta + \frac{\ddot w}{w} &= -\tT{r}{r} \\
		{\alpha'}^2 - \dot \alpha^2 + \eta'' - \ddot \eta  &= \tT{\phi}{\phi} \\
		2 \alpha' \dot \alpha -  \frac{w'}{w} \dot \eta - \frac{\dot w}{w} \eta' + \frac{\dot w'}{w} &= \tT{t}{r} \; ,
	\end{align}
\end{subequations}
where we defined
\begin{equation}
	\tT{A}{B} := \re^{2(\eta - \alpha)}\, T^A_{\phantom{A}B}
\end{equation}
with $ T^A_{\phantom{A}B} $ being the energy momentum tensor.

The symmetry of the metric \eqref{eq:met_cyl_symm} allows us to identify $ \mathbf{x} \propto \boldsymbol{\partial_r} $ as the direction of wave propagation, and the corresponding mixed null tetrad is
\begin{subequations}
\label{eq:mixed_tet_cyl}
\begin{align}
	\mathbf{k} &= \frac{\re^{\alpha - \eta}}{\sqrt{2}}\left( \boldsymbol{\partial}_t + \boldsymbol{\partial}_r\right),&
	\mathbf{l} &= \frac{\re^{\alpha - \eta}}{\sqrt{2}}\left( \boldsymbol{\partial}_t - \boldsymbol{\partial}_r\right),\\
	\mathbf{m}_1 &= \re^{-\alpha} \boldsymbol{\partial}_{z} \; , & \mathbf{m}_2 &= \frac{\re^{\alpha}}{w} \boldsymbol{\partial}_{\phi} \; .
\end{align}
\end{subequations}
Again, the vector $ \re^{\alpha - \eta}\boldsymbol{\partial_t} $ is in general not tangent to a geodesic, but due to the symmetries of the metric \eqref{eq:met_cyl_symm}, the same argument as presented below equation \eqref{eq:mixed_tet_static_plane} in the static planar case applies here as well. Thus, according to the standard interpretation, the primed components of the Weyl tensor ($ \Psi' $ and $ \Omega' $) evaluated in the frame \eqref{eq:mixed_tet_cyl} should correspond to \textit{outgoing} waves, and the unprimed ones ($ \Psi $ and $ \Omega $) to \textit{incoming} waves.

\subsection{Static cylinder\label{sec:stat_cyl}}

Let us first discuss the static solution, i.e.\ all metric functions only depend on $ r $, choosing a hollow cylinder of matter located at $ r = r_0 $ as a source:
\begin{subequations}
\begin{align}
\label{eq:emt_static}
	T^{A}_{\phantom{A}B} & = \text{diag}\left(-\rho, 0, p_z, p_{\phi}\right) \frac{1}{w} \delta(r-r_0) \; .
\end{align}
\end{subequations}
The metric outside such a static cylinder was first derived by Levi-Civita \cite{Levi:1919} (see also, e.g.\ \cite{Thorne:1965}). Like in the planar case, the full solution of Einstein's field equations can be obtained by first solving the homogeneous equations inside and outside the cylinder, and then matching the solutions such that the metric is continuous across the cylinder, but the first $ r $ derivatives are discontinuous in order for the second $ r $ derivatives on the left-hand side to produce the correct $ \delta $-like contribution \eqref{eq:emt_static} on the right-hand side. By further implementing the regularity conditions \eqref{eq:symmetry_axis} and \eqref{eq:regularity_cond}, we arrive at
\begin{subequations}
\label{eq:static_sol_pphi}
\begin{align}
	w(r) & =
	\begin{cases}
		r & (r < r_0) \\
		r_0 + w_1(r - r_0) & (r > r_0)
	\end{cases}\\
	\alpha(r) & =
	\begin{cases}
		0 & (r < r_0) \\
		\alpha_1 \ln\left[w(r)/r_0\right] & (r > r_0)
	\end{cases}\\
	\eta(r) & =
	\begin{cases}
		0 & (r < r_0) \\
		\alpha_1^2 \ln\left[w(r)/r_0\right] & (r > r_0) \; .
	\end{cases}
\end{align}
\end{subequations}
Inside the cylinder the metric is just that of flat Minkowski space, and outside it has the form
\begin{multline}
\label{eq:met_static_sol_pphi}
	\rd s^2 = \left(\frac{w(r)}{r_0}\right)^{2 \alpha_1(\alpha_1-1)} \left( -\rd t^2 + \rd r^2 \right) \\
	+ \left(\frac{w(r)}{r_0}\right)^{2 \alpha_1} \rd z^2 + \left(\frac{w(r)}{r_0}\right)^{2(1-\alpha_1)} r_0^2  \rd\phi^2 \; .
\end{multline}
From the junction conditions across $ r=r_0 $ it follows that the constants $ \alpha_1 $ and $ w_1 $ are related to the source localized on the cylinder surface by
\begin{align}
\label{eq:mc_cyl_static}
	w_1 = 1 - \rho \; , && \alpha_1 w_1 = -\frac{1}{2} \left( \rho + p_z - p_\phi \right),
\end{align}
as well as the following relation between the energy-momentum components:
\begin{equation}
\label{eq:eos_cyl_static}
	4 \left(1 - \rho \right) p_\phi = \left( \rho + p_z - p_\phi \right)^2 ,
\end{equation}
Physically, this equation of state again originates from the requirement for the source to be in hydrostatic equilibrium, since otherwise the solution would not be static. Note that for $ p_\phi = 0 $ these relations imply the equation of state $ p = - \rho $ as well as $ \alpha_1 = 0 $, and the solution becomes the geometry of a cosmic string \cite{Vilenkin:1981zs, Hiscock:1985uc}, which only produces a deficit angle in the outside geometry, and has a vanishing Weyl tensor for $ r>r_0 $. If $ p_\phi \neq 0 $, however, the parameter $ \alpha_1 $ can in principle take any value, leading to an outside geometry that is not Riemann flat. The Weyl components for the solution \eqref{eq:static_sol_pphi} become, for $ r > r_0 $,
\begin{subequations}
\label{eq:weyl_cyl_static}
\begin{align}
		\Phi &= \Phi_0 \, \re^{2(\alpha-\eta)}\: , \quad \Phi_{ij} = -\frac{1}{2} \Phi_0 \, \re^{2(\alpha-\eta)} \delta_{ij}\\
		\Omega_{ij} & = \Omega'_{ij} = \Omega \, \re^{2(\alpha-\eta)}
		\begin{pmatrix}
		1 & 0 \\
		0 & -1
		\end{pmatrix}\\
		\intertext{with}
		\Omega & := \frac{1}{2}\alpha_1 (\alpha_1 - 1) (2\alpha_1 - 1) w_1^2\, w(r)^{-2} \; ,\\
		\Phi_0 & := -\alpha_1 (\alpha_1 - 1) w_1^2\, w(r)^{-2} \; .
\end{align}
\end{subequations}
Incidentally, this shows that the geometry is always asymptotically flat for $ r \to \infty $ because
\begin{equation}
	\re^{2(\alpha-\eta)} w(r)^{-2} = r_0^2 w(r)^{-2 \left(\alpha_1^2 - \alpha_1 + 1 \right)}
\end{equation}
and
\begin{equation}
	2 \left(\alpha_1^2 - \alpha_1 + 1 \right) > \frac{3}{2} \quad \forall \alpha_1 \; ,
\end{equation}
and we assume $ w_1 > 0 $ in order to avoid a second axis outside the cylinder, where $ w(r) $ would be zero. Moreover, all Weyl components have the same falloff behavior which is $ \sim r^{-2} $ for $ \Phi_0 $ and $ \Omega $ .

The problem with the standard interpretation is now manifest: 
if the standard interpretation of the individual components were correct, 
we would conclude that (for $ \alpha_1 \neq 0, 1, 1/2 $) there are
incoming and outgoing waves present in this solution.
The geometry, however, is in fact \textit{static}, so we know for sure that there are actually no waves at all.
This shows that the $ \Omega $ parts of the Weyl tensor are not only 
due to gravitational wave components, but also due to static, i.e.\ Newton-like field components. 
There is no contradiction to the discussion of the geodesic deviation 
in section \ref{sec:interpret_weyl}, because the tidal forces in the $ z $ and $ \phi $ directions, acting on a freely falling observer in the cylindrical geometry \eqref{eq:met_static_sol_pphi}, 
will in general not be equal, and, therefore, can produce an elliptical deformation of test particles 
precisely in the same way as discussed in section \ref{sec:interpret_weyl}. 
As a result, the $ \Omega $ parts of the Weyl tensor 
cannot be used to extract the purely wavelike content of the space-time geometry. 

It should be noted that this also provides a counterexample for a more recent suggestion of a ``radiation scalar'' \cite{Beetle:2002}, which was defined as the product of the complex Weyl scalars $ \Psi_0 $ and $ \Psi_4 $, in a tetrad frame in which $ \Psi_1 $ and $ \Psi_3 $ are zero \footnote{The relation between the complex Weyl scalars, usually used in four space-time dimensions, and the real ones that we use, can be found in \cite{Podolsky:2012he}}. Using our variables, the latter requirement translates to $ \Psi = \Psi' = 0 $, and is thus satisfied by our static solution. Furthermore, in this solution the radiation scalar becomes $ \Omega^2 $, which is in general nonvanishing. This shows that the following claim in \cite{Beetle:2002},
\begin{quote}
	``[the radiation scalar] vanishes in regions of space-time which can be said unambiguously to contain no gravitational radiation'',
\end{quote}
is not true.

However, one might still hope that even though the interpretation of the $ \Omega $ terms fails for static geometries, it could still be valid for pure wave solutions, and in particular the distinction between incoming and outgoing waves could still be rigorously made for those cases. Unfortunately, this is also not true, as we will show in the next section.

\subsection{Einstein-Rosen waves}
To this end, consider cylindrically symmetric gravitational waves, first discovered by Einstein and Rosen \cite{EinsteinRosen1937} (see also, e.g.\ \cite{Marder1958}). The metric for this class of solutions is obtained from \eqref{eq:met_cyl_symm} by setting $ w=r $:
\begin{equation}
\label{eq:met_ER}
	\rd s^2 = \re^{2(\eta - \alpha)} \left( -\rd t^2 + \rd r^2 \right) + \re^{2\alpha} \rd z^2 + \re^{-2\alpha} r^2 \rd\phi^2 \; .
\end{equation}
We refrain from specifying the energy-momentum tensor that would give rise to the cylindrical waves and simply assume that there is some time-dependent source distributed in accordance with the cylindrical symmetry over a bounded region around the axis at $ r=0 $. Thus, we only consider the vacuum Einstein equations outside this region, which take the simple form
\begin{subequations}
\label{eq:EFE_ER}
	\begin{gather}
	\label{eq:wave_eq_alpha}
		\alpha'' +  \frac{\alpha'}{r} - \ddot \alpha = 0 \; , \\
	\label{eq:EFE_ER_eta}
		\eta' = r \left( {\alpha'}^2 + \dot \alpha^2 \right), \quad \dot \eta = 2r \alpha' \dot \alpha \; .
	\end{gather}
\end{subequations}
The first equation is nothing but the linear cylindrical wave equation in flat space, and given any solution $ \alpha $ of this equation, the other two --- which are consistent on account of the first one --- can simply be integrated to obtain $ \eta $.

\begin{figure*}
	\includegraphics[width=0.49\textwidth]{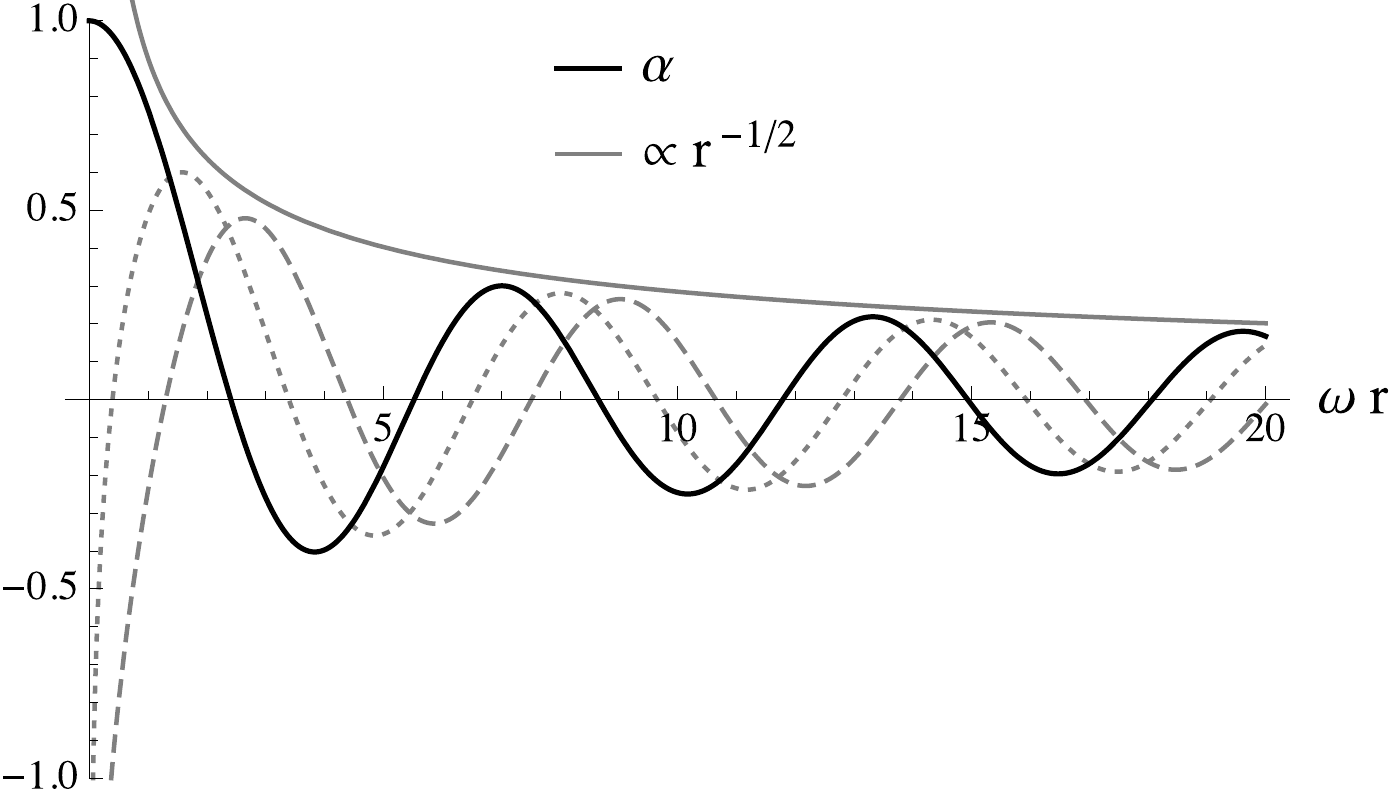}\hfill
	\includegraphics[width=0.49\textwidth]{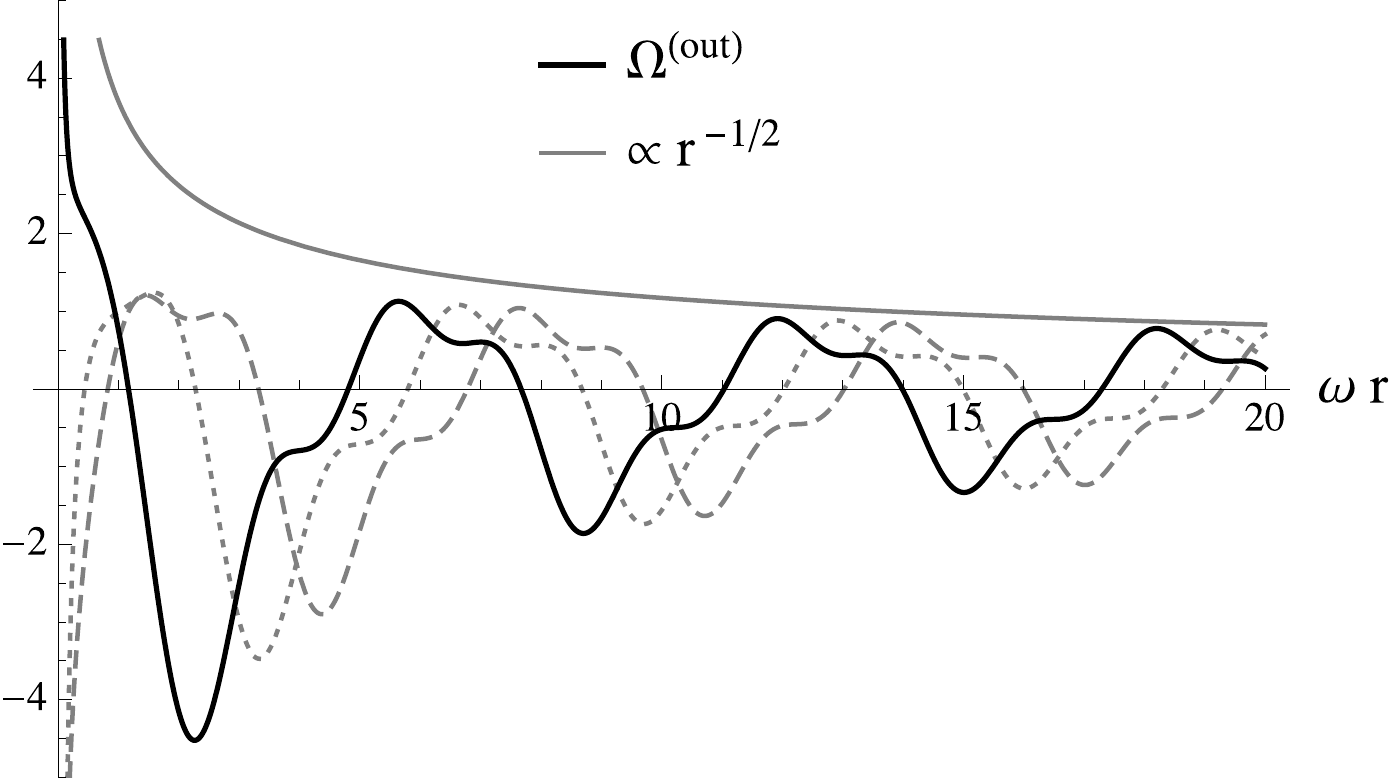}
	\includegraphics[width=0.49\textwidth]{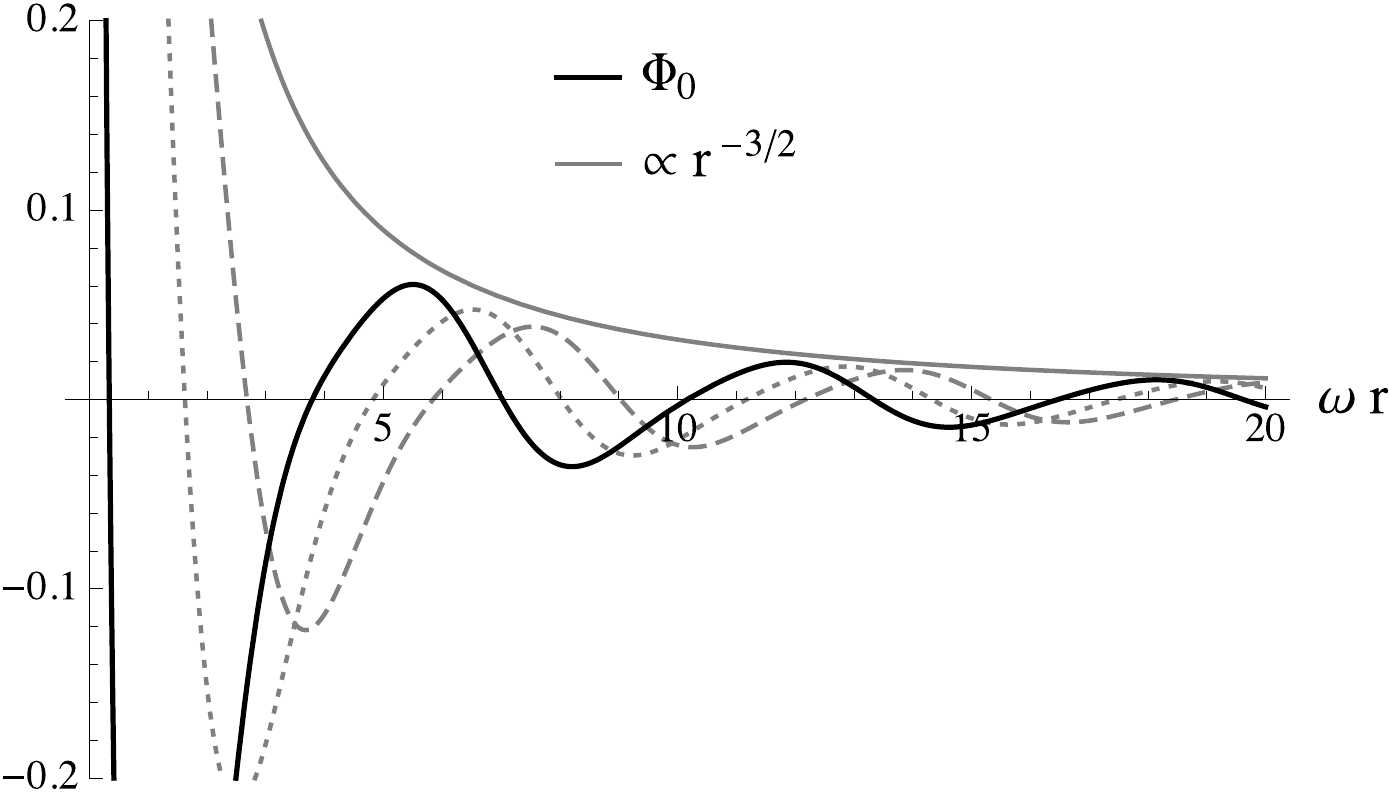}\hfill
	\includegraphics[width=0.49\textwidth]{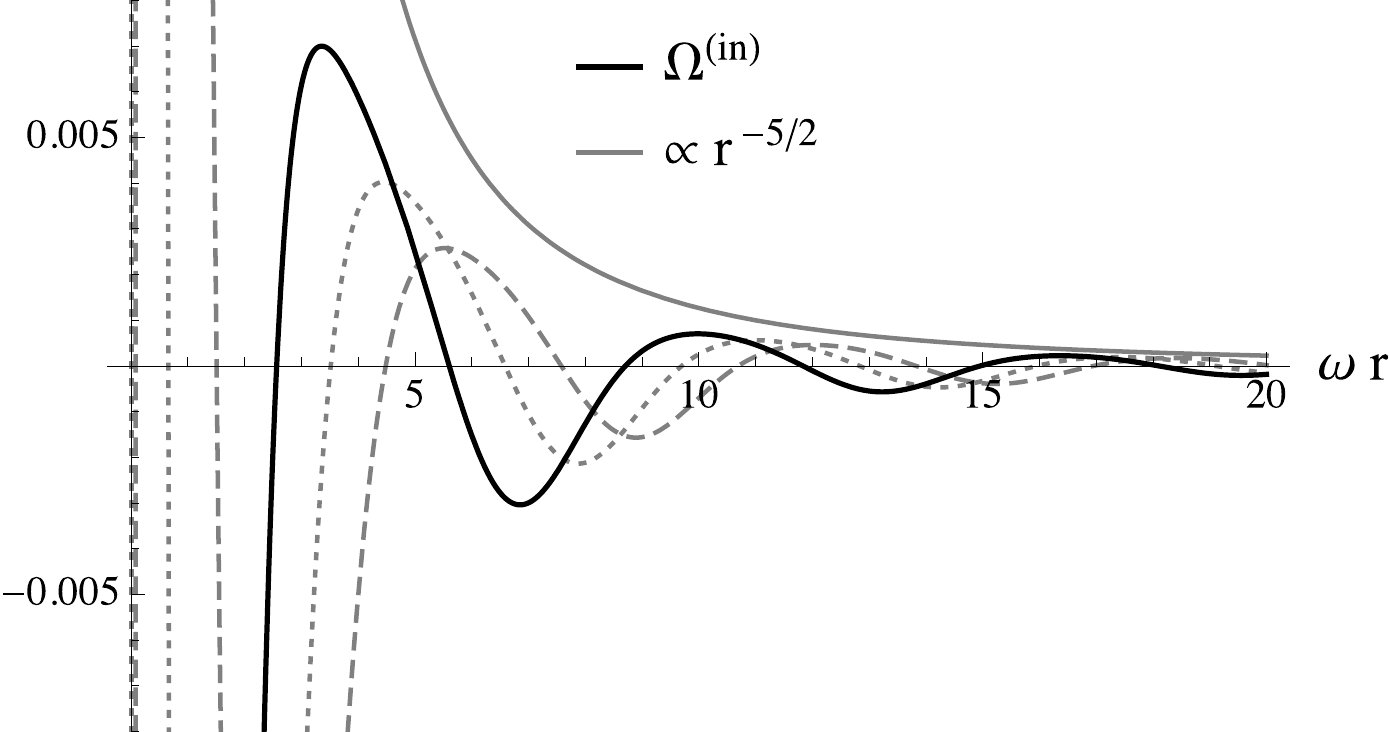}
	\caption{Plots of the metric function $ \alpha $ and all the independent components of the Weyl tensor for the outgoing Einstein-Rosen wave solution \eqref{eq:alpha_outgoing}. The solid black lines correspond to $ t=0 $, whereas the dotted and dashed lines are evaluated at times $ t = 1 / \omega $ and $ t = 2 / \omega $, respectively. Evidently, all functions behave like waves traveling outwards. All the terms have different falloff behavior with $ r $, $ \Oout $ being the only component with the same behavior $ \propto 1/\sqrt{r} $ as a linear cylindrical wave, but they are all nonvanishing.}
	\label{fig:ER_outWave}
\end{figure*}

Since this system is so simple, but still represents exact solutions of Einstein's field equations with a clear physical interpretation, it is a perfect candidate to test the standard interpretation of the Weyl components. For the metric \eqref{eq:met_ER}, the nonzero, independent components of the Weyl tensor become, after some simplifications using \eqref{eq:EFE_ER},
\begin{subequations}
\label{eq:weyl_comp_ER}
\begin{align}
	\Omega_{ij} & = \Oin\; \re^{2(\alpha-\eta)}
		\begin{pmatrix}
		1 & 0 \\
		0 & -1
		\end{pmatrix},\\
	\Phi & = \Phi_0\; \re^{2(\alpha-\eta)} \; , \\
	\Omega'_{ij} & = \Oout\; \re^{2(\alpha-\eta)}
		\begin{pmatrix}
		1 & 0 \\
		0 & -1
		\end{pmatrix},\\
	\intertext{with}
	\begin{split}
	\Oin & := -\frac{1}{2}\Bigl[ \bigl(3 - 2 r \left(\alpha' + \dot\alpha \right) \bigr) \left(\alpha' + \dot\alpha \right)^2 \\
	& \qquad\qquad + \alpha'' + 2 \dot\alpha' + \ddot\alpha \Bigr],
	\end{split}\\
	\Phi_0 & := \left(\frac{1}{r} - \alpha' \right) \alpha' + {\dot\alpha}^2 \; , \\
	\begin{split}
	\Oout & := -\frac{1}{2}\Bigl[ \bigl(3 - 2 r \left(\alpha' - \dot\alpha \right) \bigr) \left(\alpha' - \dot\alpha \right)^2 \\
	& \qquad\qquad + \alpha'' - 2 \dot\alpha' + \ddot\alpha \Bigr].
	\end{split}
\end{align}
\end{subequations}
Now consider, for example, a solution of \eqref{eq:wave_eq_alpha} corresponding to a purely outgoing wave with frequency $ \omega $:
\renewcommand{\Re}{\operatorname{Re}}
\begin{equation}
\label{eq:alpha_outgoing}
\begin{split}
	\alpha(t, r) &= \Re\left[\re^{-i \omega t} H_0^{(1)}(\omega r)\right] \\
	&= \cos(\omega t) J_0(\omega r) + \sin(\omega t) Y_0(\omega r) \; ,
\end{split}
\end{equation}
where $ H_0^{(1)} $ denotes the Hankel function of the first kind, and $ J_0 $ and $ Y_0 $ are the Bessel functions of the first and second kind, respectively. Plugging \eqref{eq:alpha_outgoing} into \eqref{eq:weyl_comp_ER} yields some complicated expressions for the three components, containing trigonometric and Bessel functions, the explicit form of which is not of great interest. The main point is that they are all nonvanishing, even though their magnitude and falloff behavior with $ r $ is different; cf.\ Figure \ref{fig:ER_outWave}. Furthermore, they all behave like outgoing waves, in accordance with our choice \eqref{eq:alpha_outgoing}, confirming that the geometry only contains outgoing wave components.

The observation that the component $ \Phi $ is nonzero is not too disturbing, because an outgoing wave could also induce Newton-like contributions --- even though the solution \eqref{eq:alpha_outgoing} does not contain a static, Newton-like part $ \propto \ln(r) $. But the fact that $ \Oin $ is nonzero, and furthermore also behaves like an outgoing wave, clearly shows that the interpretation of $ \Oin $ as an incoming gravitational wave is not correct.

Even though the interpretation that $ \Oin $ corresponds to incoming gravitational waves is wrong, there is still a difference between the different Weyl components: they all have a distinct falloff behavior with $ r \to \infty $. Only $ \Oout $ falls asymptotically off as a spatially two-dimensional wave $ \sim r^{-1/2} $, whereas all other components fall off faster. So in the case of pure cylindrical waves, the interpretation still works asymptotically far away from the source. This can physically be understood because for $ r \to \infty $ cylindrical waves look like plane waves, for which we saw that the decomposition into incoming and outgoing wave components is successful.

This might be considered a window of opportunity, indicating that the standard interpretation
of the Weyl components might in general still hold asymptotically far away from the source. However, for the static solution discussed in section \ref{sec:stat_cyl}, all the components have exactly the same falloff behavior\footnote{As discussed below equation \eqref{eq:mixed_tet_static_plane}, the different Weyl components could get overall factors if one took the true geodesic tetrad, which could change the falloff behavior. But these factors come with powers of the boost weight of the corresponding components, and so it is not possible that both $ \Omega $ and $ \Omega' $ could both become suppressed relative to $ \Phi $ for the static cylindrically symmetric solution. If one of them is suppressed, the other one will always be enhanced.}; cf.\ equation \eqref{eq:weyl_cyl_static}. Thus, it is \emph{not true} in general that the decomposition into Newton- and wavelike components gets better and better in the limit $ r \to \infty $. Physically, the key difference between the pure wave and the static case is, of course, that while waves look like plane waves far away from the source, the cylinder does not look like a plane from far away.

On the other hand, as shown in section \ref{sec:stat_cyl}, the wave components that are due to such a static field always fall off like $ \Omega \sim r^{-2} $, i.e.\ faster than $ \sim r^{-1/2} $ which is the characteristic falloff behavior of the actual gravitational waves. Therefore, the correct statement is the following: in the case of a cylindrical geometry, those parts of the wave components $ \Oin $ and $ \Oout $ that fall off like $ \sim r^{-1/2} $ are due to incoming and outgoing gravitational waves, respectively. Thus, the standard interpretation still applies to cylindrically symmetric geometries in the weaker sense that the presence and distinction of incoming and outgoing radiation can be inferred from the asymptotic falloff behavior of the respective wave components of the Weyl tensor. But our result unambiguously shows that these components cannot be used as a local criterion, i.e. by evaluating them at some fixed value of $ r $, for the presence or absence of gravitational waves.

\section{Pirani's criterion\label{sec:pirani}}

In \cite{Pirani:1957}, Pirani suggested a slightly different criterion for deciding whether a given geometry contains  gravitational waves or not, depending on its Petrov type:
\begin{quote}
``At any event in empty space-time, gravitational radiation is present if the Riemann tensor is of Type II or Type III, but not if it is of Type I.''
\end{quote}
(Note that here type D and N should be understood as subclasses included in type I and II, respectively.)
The difference to the Weyl component criterion is that a Petrov type I space-time can have nonvanishing components $ \Omega $ and $ \Omega' $ \cite[chap. 4.2]{Stephani}. However, they must be equal, and so Pirani would correctly classify the static example of section \ref{sec:stat_cyl} as one without waves, because it is of type I. However, it turns out that the dynamical solution of Eintsein-Rosen waves can in fact be of type I. Actually, this example was also discussed by Pirani in \cite{Pirani:1957}, claiming that it is of type II and thus in accordance with his definition. But this statement is false: even though they can be of type II, they can also be of type I \cite[p. 352]{Stephani}, as is for example the case for our outgoing wave solution. Therefore, Pirani's diagnostic tool also does not work in general.

\section{Thorne's C-energy\label{sec:thorne}}

For whole-cylindrically symmetric space-time geometries, like the ones discussed in section \ref{sec:interpret_fails}, Thorne was able to define a physically well-motivated notion of local energy, the so called ``C-energy'' \cite{Thorne:1965}. It is given by a covariantly conserved four-vector $ \boldsymbol{P} $, which can be viewed as the flux vector of gravitational energy. For an (accelerated) observer localized at constant spatial coordinates, $ P^{t} $ is thus the energy density, and $ P^{r} $ corresponds to the flux of energy in radial direction. Therefore, $ P^{r} $ could be used as a diagnostic tool, which should be nonzero only if gravitational waves are present, and the sign of which should determine the direction of propagation.

Let us, for convenience, briefly recall the definition from \cite{Thorne:1965}. The C-energy flux vector $ \boldsymbol{P} $ is defined as
\begin{equation}
	P^{A} := \frac{\epsilon^{ABCD}}{\sqrt{-g}} \left( \partial_B E \right) \frac{\xi_{(z)C}}{|\boldsymbol{\xi}_{(z)}|^2} \frac{\xi_{(\phi)D}}{|\boldsymbol{\xi}_{(\phi)}|^2} \; ,
\end{equation}
where $ \epsilon^{ABCD} = +1 $ when $ (ABCD) $ are even permutations of $ (trz\phi) $, $ \boldsymbol{\xi}_{(z)} $ and $ \boldsymbol{\xi}_{(\phi)} $ are the two Killing vectors of the whole cylindrical geometry (normalized such that  $ |\boldsymbol{\xi}_{(z)}| = 1 $ on the symmetry axis when there is no gravitational radiation present, and $ |\boldsymbol{\xi}_{(\phi)}| $ measures the proper circumference around the axis), and $ E $ denotes the ``potential function'' for C-energy:
\begin{equation}
E := -\pi \ln\left[ \frac{ \partial_A\left( |\boldsymbol{\xi}_{(z)}| |\boldsymbol{\xi}_{(\phi)}| \right) \partial^A\left( |\boldsymbol{\xi}_{(z)}| |\boldsymbol{\xi}_{(\phi)}| \right)}{4\pi^2  |\boldsymbol{\xi}_{(\phi)}|^2} \right].
\end{equation}

In the coordinates \eqref{eq:met_cyl_symm},
the Killing vectors have the components
\begin{equation}
	\xi_{(z)}^A = h_z \delta^A_z, \quad 	\xi_{(\phi)}^A = 2\pi \delta^A_\phi \; ,
\end{equation}
where $ h_z $ is some positive constant [which is fixed by requiring the proper normalization of $ \boldsymbol{\xi}_{(z)} $]. The potential function then becomes
\begin{equation}
\label{eq:C_energy_pot}
	E = 2\pi \left[ \eta - \frac{1}{2} \ln\left( {w'}^2 - {\dot w}^2 \right) \right] ,
\end{equation}
and the nonvanishing components of the flux vector are
\begin{subequations}
\label{eq:C_energy_flux}
\begin{align}
	P^{t} & = + \frac{1}{2\pi h_z} \frac{\re^{2(\alpha - \eta)}}{w} E' \; ,\\
\label{eq:C_energy_flux_r}
	P^{r} & = - \frac{1}{2\pi h_z} \frac{\re^{2(\alpha - \eta)}}{w} \dot E \; .
\end{align}
\end{subequations}
Inserting \eqref{eq:C_energy_pot} in \eqref{eq:C_energy_flux}, and using the Einstein field equations \eqref{eq:EFE_cyl} in vacuum, the flux vector can be written as
\begin{subequations}
\begin{align}
	P^{t} & = C \frac{w' (\dot\alpha^2 + {\alpha'}^2) - 2 \dot w \dot\alpha \alpha'}{{w'}^2 - \dot w^2} \; ,\\
	P^{r} & = C \frac{\dot w (\dot\alpha^2 + {\alpha'}^2) - 2 w' \dot\alpha \alpha'}{{w'}^2 - \dot w^2} \; ,
\end{align}
\end{subequations}
where $ C $ denotes the manifestly positive --- and thus for our purposes not important --- factor $ C := \exp[2(\alpha - \eta)] / h_z $.

From \eqref{eq:C_energy_flux_r} it follows immediately that the energy flux $ P^r $ indeed vanishes for the static solution \eqref{eq:static_sol_pphi}, which has the only nonzero component
\begin{equation}
	P^{t} = C \alpha_1^2 w_1 w(r)^{-2} \; .
\end{equation}
So this time the static, Newton-like character of the Weyl tensor is correctly reproduced. Furthermore, $ P^t $ will be positive, as long as $ w_1 > 0 $, which according to \eqref{eq:mc_cyl_static} and \eqref{eq:eos_cyl_static} is equivalent to $ p_\phi > 0 $ and also to $ \rho < 1 $, and which is necessary for the geometry outside the cylinder to be regular. The positivity of $ P^t $ is of course necessary for it to be interpreted as an energy density.

\begin{figure*}
	\includegraphics[width=0.49\textwidth]{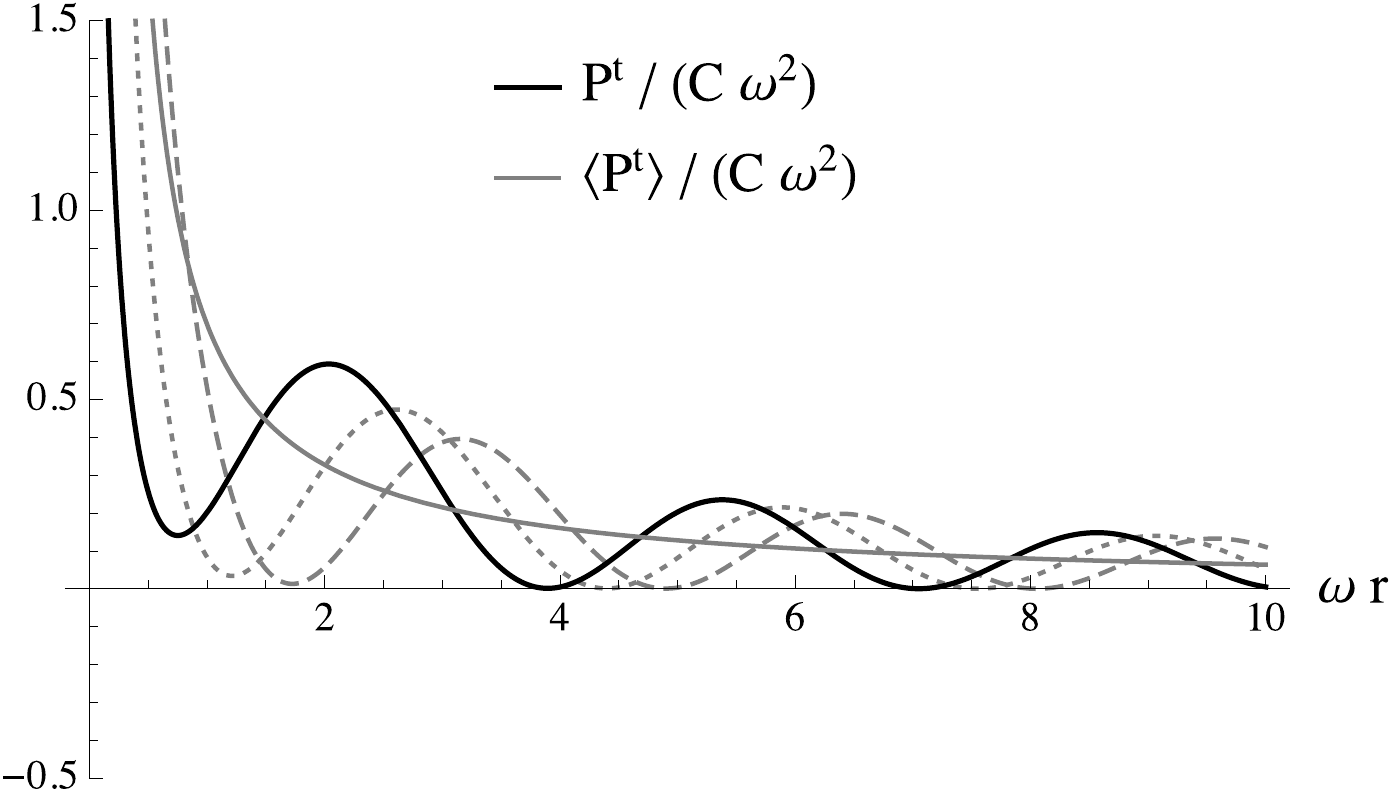}\hfill
	\includegraphics[width=0.49\textwidth]{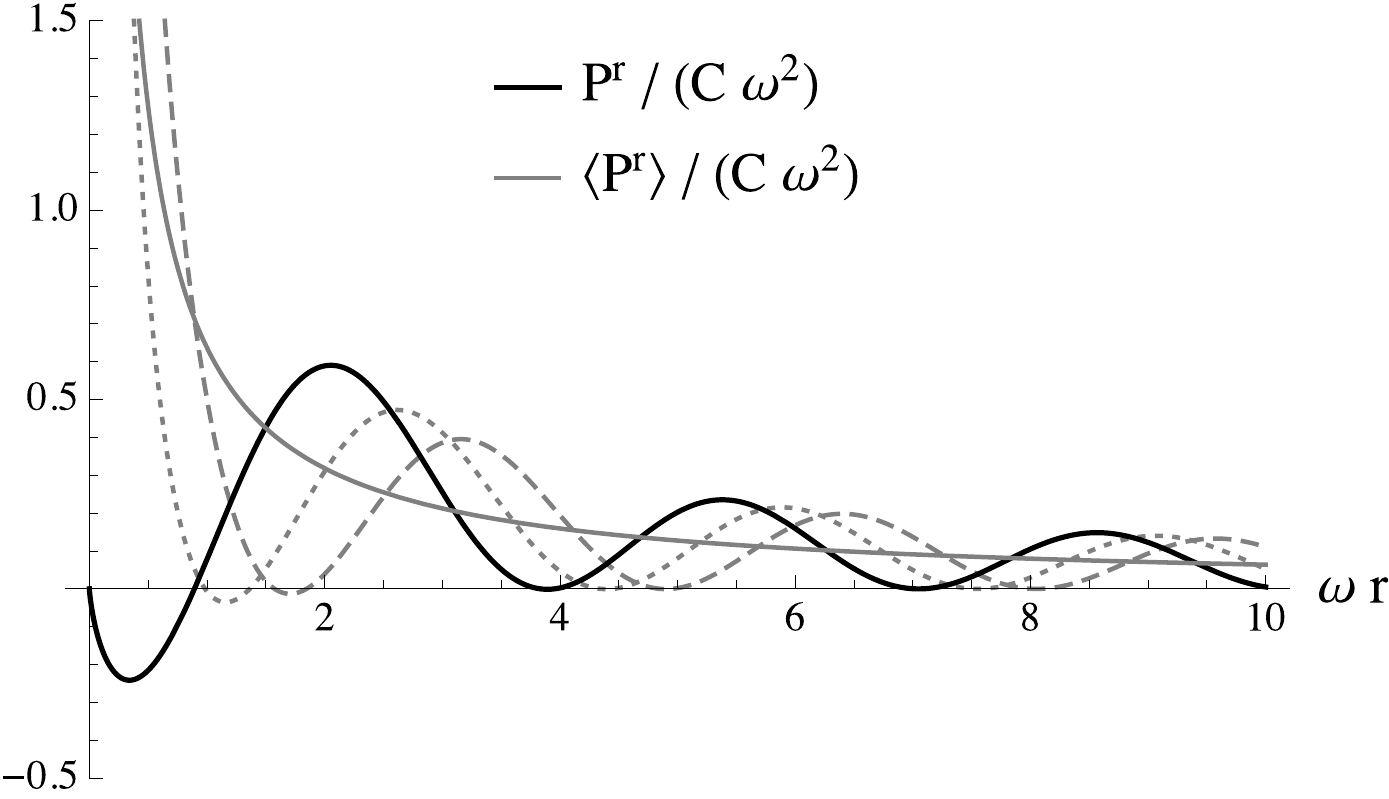}
	\caption{Plots of the nonvanishing components of the C-energy flux vector for the outgoing Einstein-Rosen wave solution \eqref{eq:alpha_outgoing}. The solid black lines correspond to $ t=0 $, whereas the dotted and dashed lines are evaluated at times $ t = 1 / (2\omega)$ and $ t = 1 / \omega $, respectively. The gray lines are the corresponding time-averaged quantities, which (asymptotically) fall off like $ 1/r $.}
	\label{fig:CEnergy}
\end{figure*}

For the Einstein-Rosen geometry \eqref{eq:met_ER}, the flux vector in vacuum becomes
\begin{subequations}
\begin{align}
	P^{t} & = C \left( \dot\alpha^2 + {\alpha'}^2 \right) \; ,\\
	P^{r} & = -2C \dot\alpha \alpha' \; ,
\end{align}
\end{subequations}
which for the outgoing wave \eqref{eq:alpha_outgoing} gives
\begin{subequations}
\begin{align}
\begin{split}
	P^t & = C \omega^2 \left\{\bigl[J_1(\omega r) \cos(\omega t) + Y_1(\omega r) \sin(\omega t) \bigr]^2 \right. \\
		& \left. \quad + \bigl[Y_0(\omega r) \cos (\omega t) - J_0(\omega r) \sin(\omega t) \bigr]^2 \right\}, 
\end{split}\\
\begin{split}
	P^r & = 2 C  \omega^2 \bigl[Y_0(\omega r) \cos(\omega t) - J_0(\omega r) \sin(\omega t)\bigr] \\
		& \quad \times \bigl[J_1(\omega r) \cos(\omega t) + Y_1(\omega r) \sin(\omega t)\bigr].
\end{split}
\end{align}
\end{subequations}
First of all, note that $ P^t $ is manifestly non-negative, as is required for its interpretation as an energy density. Furthermore, the flux in $ r $ direction does not vanish, in agreement with the presence of gravitational waves. As can be seen from the plot in Figure \ref{fig:CEnergy}, it is \textit{not} strictly positive, though; in the region close to the axis at $ r=0 $, it becomes negative. This effect, however, disappears if the flux averaged over one period $ T \equiv 2\pi/\omega $ is considered:
\begin{equation}
\langle P^r \rangle := \frac{1}{T} \int_0^T P^r \rd t = \frac{2 C \omega}{\pi r}
\end{equation}
This has exactly the form expected for a stationary flux of outgoing cylindrical waves. Purely incoming Einstein-Rosen waves, $ \alpha \sim \Re[ \re^{-i\omega t} H_0^{(2)}(\omega r) ] $ instead of \eqref{eq:alpha_outgoing}, yield the same result, but with opposite sign.

Let us emphasize that, on the other hand, $ \langle\Oin\rangle \neq 0 $ and so this averaging cannot be used to circumvent the failure of the standard Weyl interpretation.

As a result, Thorne's C-energy passes all tests in our two examples, where the standard interpretation of the Weyl tensor fails, apart from the subtlety that the radial flux $ P^r $ can locally (both in space and time) have the opposite sign than the actual direction of wave propagation. This could be due the well-known fact \cite{Hadamard} that waves in two spatial dimensions do not obey the principle of Huygens and Fresnel, meaning that signals do not propagate exclusively with the speed of light, but also slower. (This can directly be seen from the fact that the retarded Green's function for the wave operator in $ 2+1 $ dimensions has support not only \emph{on} the backward light cone, but also \emph{inside}.) A possible interpretation is based on backscattering processes taking place, because if we still demand propagation to mean propagation with the fundamental velocity, then an outgoing wave in two spatial dimensions must be viewed as the superposition of a part propagating outwards and some (smaller) part that propagates inwards. Provided the C-energy flux measures energy propagating with the speed of light, backscattering processes could explain the unexpected behavior of $ P^r $. Only for large distances from the axis, where the waves asymptotically become one-dimensional, the backscattering effect becomes negligible. This would also explain why the ``wrong'' sign of $ P^r $ is only found close to the axis, where backscattering is still efficient.

This observation suggests that the same effect might also be responsible for the failure of the Weyl-criterion when applied to cylindrical waves. In this case there is also a further, independent argument: if $ \Omega $ really corresponded to the purely incoming gravitational wave component, then this term, when set equal to zero, would constitute a \textit{local} outgoing-wave condition. However, no such local condition exists for linear cylindrical waves, because the wave operator does not factorize in two spatial dimensions --- unlike in one or three, where
\begin{subequations}
\begin{align}
	-\partial_t^2 + \Delta_{(1)} & = \left( -\partial_t + \partial_r \right) \left( \partial_t + \partial_r \right), \\
	-\partial_t^2 + \Delta_{(3)} & = \left( -\partial_t + \frac{1}{r} + \partial_r \right) \left( \partial_t + \frac{1}{r} + \partial_r \right).
\end{align}
\end{subequations}
It would thus be quite surprising if the much more complicated nonlinear wave phenomena of general relativity would in general allow for such a local condition\footnote{The only way this could in principle still be possible for Einstein-Rosen waves is that the criterion can involve not only $ \alpha $, but also $ \eta $, which when expressed in terms of $ \alpha $ by means of \eqref{eq:EFE_ER_eta} becomes nonlocal.}. Note that this argument is also supported by the observation that the standard interpretation seems to work for plane and spherical waves, but fails for cylindrical waves.

On the other hand, this still does not explain why the $ \Omega $ terms are nonzero in the static cylindrically symmetric solution.

\section{Conclusion\label{sec:conclusion}}

The main scientific objective of this work was to prove that the standard interpretation
of the Weyl tensor \cite{Szekeres:1965ux, Stephani, Podolsky:2012he} does not hold 
in general. 
This has been achieved by investigating explicitly two exact vacuum solutions of Einstein's field equations with whole-cylinder symmetry: first, a static solution has been shown to admit nonvanishing wavelike components according to the standard interpretation, which does make sense. In addition, this example invalidates a more recent suggestion for a local criterion indicating the presence of gravitational waves \cite{Beetle:2002}. Second, purely outgoing Einstein-Rosen waves were considered, and it was shown that according to the standard interpretation, they have not only nonzero outgoing wave- and Newton-like components, but also \textit{incoming} wavelike components. Moreover, they are of Petrov Type I, which also shows that a further instrument to diagnose 
the presence of wavelike components put forward by Pirani \cite{Pirani:1957}
does not work.

The only way in which the standard interpretation can still be used in the case of cylindrical symmetry is by considering only those parts of the Weyl components which have the correct asymptotic falloff behavior as $ r \to \infty $: $ \sim r^{-2} $ for the Newton-like component $ \Phi_0 $, and $ \sim r^{-1/2} $ for the transverse wavelike components $ \Oin $ and $ \Oout $. But it can not be used as a spatially \emph{local} criterion.

Furthermore, we showed that the standard interpretation \emph{does} work for plane wave and spherical wave geometries. Based on these findings, we argued that the failure of the wave interpretation might be due to the more fundamental inability to locally distinguish between incoming and outgoing wave components for cylindrical waves, which is even true for
the weak field limit on a Minkowski background. This can be seen by the fact that the d'Alembert operator does not factorize in two spatial dimensions, whereas this factorization property is granted in one or three spatial dimensions.

Therefore, the standard interpretation of the Weyl tensor cannot be trusted when
space-time geometries with whole-cylinder symmetry are being investigated. Instead, for those systems, Thorne's local energy concept, called C-energy, is an appropriate tool to diagnose the presence or absence of gravitational waves. It is, however, not possible to locally distinguish between incoming and outgoing waves, for the very same reason stated before. At least, Thorne's concept works for oscillatory waves when the C-energy flux 
averaged over a time period is considered. This gives a spatially local criterion 
with the correct interpretation. 

Note that the above explanation actually addresses only the distinction between incoming and outgoing wavelike components, but does not resolve the problem that a static solution can admit nonzero wavelike contributions. However, it was also shown that the wave components in fact \textit{do} vanish for \textit{static} plane-symmetric and spherically symmetric geometries. This gives prominence to the cylinder symmetry as the cause 
for the standard interpretation to fail.

Finally, let us stress that our results do not invalidate the Gedankenexperiment to measure components of the Weyl tensor by observing the relative acceleration of nearby freely falling test particles. Our work rather shows that it is in general not possible to classify these components as Newton-like, or
as incoming or outgoing gravitational wave components.

\begin{acknowledgements}
The authors would like to thank Stanley Deser, Felix Berkhahn, Dennis Schimmel and Tehseen Rug for inspiring discussions. The work of SH was supported by the DFG cluster of excellence `Origin and Structure of the Universe' and by TRR 33 `The Dark Universe'. The work of FN and RS was supported by the DFG cluster of excellence `Origin and Structure of the Universe'.

\end{acknowledgements}

\newpage
\bibliography{references}

\end{document}